\documentclass[12pt,epsf]{article}
\input epsf.tex
\def\be{\begin{equation}} \def\ee{\end{equation}}
\def\bi{\begin{itemize}} \def\ei{\end{itemize}}
\def\bea{\begin{eqnarray}} \def\eea{\end{eqnarray}} \def\ba{\begin{array}}
\def\ea{\end{array}} \def\ben{\begin{enumerate}} \def\een{\end{enumerate}}

\newcommand{\eqn}[1]{(\ref{#1})}

\newcommand{\plb}[3]{Phys. Lett. {\bf B#1} ({#2}) {#3}}
\newcommand{\prd}[3]{Phys. Rev. {\bf D#1} ({#2}) {#3}}
\newcommand{\hepth}[1]{{\tt [arXiv:{#1}[hep-th]]}}

\def\br{\nonumber\\}
\def\ud{\underline}
\def\g{\gamma}

\begin{document}
{}~
\hfill\vbox{\hbox{hep-th/1203.nnnn} 
\hbox{\today}}\break

\vskip 3.5cm
\centerline{\large \bf
Lifshitz/Schr\"odinger D$p$-branes and  dynamical exponents }
\vskip .5cm

\vspace*{.5cm}

\centerline{\bf  
Harvendra Singh}

\vspace*{.5cm}
\centerline{ \it  Theory Division} 
\centerline{ \it  Saha Institute of Nuclear Physics} 
\centerline{ \it  1/AF Bidhannagar, Kolkata 700064, India}
\vspace*{.25cm}

\vspace*{.5cm}

\vskip.5cm

\vskip1cm

\centerline{\bf Abstract} \bigskip

We extend our earlier
study of special double limits of  `boosted' $AdS_5$ black hole
solutions to include all black D$p$-branes of type II strings. 
We find that  Lifshitz solutions can be 
obtained in generality, with varied dynamical exponents,
 by employing these  limits. We then study such
double limits for `boosted'  D$p$-brane bubble solutions and  
find that the resulting non-relativistic solutions
instead describe  
Schr\"odinger like spacetimes, having varied dynamical exponents. 
We get a simple map between these Lifshitz \& Schr\"odinger solutions and
a relationship between two types of dynamical exponents.
We also discuss about the singularities of the Lifshitz solutions
and an intriguing thermodynamic duality.

\vfill 
\eject

\baselineskip=16.2pt


\section{Introduction}

The holographic studies involving non-relativistic 
string theory backgrounds 
have received good attention recently, both from the  point of view of
uncovering AdS/CFT holographic principle, as well as to
understand  some quantum critical phenomena 
in  strongly correlated condensed matter systems. 
The condensed matter systems by nature 
are usually nonrelativistic but in some  systems near some
critical point the system may appear to show exotic scaling behaviour. 
These could be systems, for example 
atomic gases at ultra low temperatures, or
fermions at unitarity \cite{son, bala}.
Such systems are known to exhibit  Schr\"odinger or 
Lifshitz like behaviour. 
 
     There have been namely two types of
non-relativistic (Galilean) string backgrounds, 
exhibiting asymmetric space and time   
scaling symmetry, which  have been a subject 
of several studies recently \cite{son}-\cite{Horowitz:2011gh}. 
The ones which exhibit  Schr\"odinger symmetries \cite{son,bala} 
are described as
\bea\label{sol1}
&&ds^2_{Sch}= \left( -
{(dx^{+})^2\over z^{2a}}  -{dx^{+}dx^{-}\over z^2}
+{dx_i^2\over z^2} \right) +{dz^2\over  z^2}  
\eea 
and those having Lifshitz type symmetries  \cite{kachru} are like
\bea\label{sol1a}
&&ds^2_{Lif}= \left( -{dt^2\over z^{2a}} 
+{dx_i^2\over z^2} \right) +{dz^2\over  z^2}.  
\eea 
In both these cases, $x^i$'s are flat spatial coordinates while 
$z$ is known as the holographic coordinate (which is a measure of the
energy scale of the boundary theory). The constant
parameter $a$ is  called the dynamical exponent of time. 
These bulk geometries are suitable to describe an
asymmetric scaling quantum phenomena at some
critical point in a nonrelativistic CFT, 
which resides on the boundary of 
such spacetimes. 
The Schr\"odinger  spacetimes can be obtained as solutions in string theory
 \cite{herzogrev,malda}. Not only this, the nonrelativistic
spaces can also be 
obtained as solutions of `massive' type IIA 
string theory  \cite{hs}. 
More recently, 
the Lifshitz-like spaces have been embedded variously in ordinary 
string theory as 
 demonstrated by \cite{bala3,donos10,Singh:2010zs}. 
These explicit examples imply that a wide class of
non-relativistic solutions can be constructed in string theory and 
 hopefully some of these
 could be engineered to explain some strongly coupled  
phenomena in respective boundary   theories.
In this work 
we shall show that a class of these solutions indeed can be obtained 
consistently 
as limiting cases of well studied  AdS D$p$ brane solutions.

We wish to extend our last study of D3-branes  where vanishing horizon
 limits of 
`boosted' $AdS_5\times S^5$ black holes give rise to the Lifshitz  solutions.
 These $a=3$ Lifshitz solutions and similar solutions in M-theory 
with fractional
dynamical exponents were studied in 
\cite{Singh:2010zs,Singh:2010cj}.  Repeating the same methods for all 
D$p$-branes here,
we obtain a wide class of Lifshitz and Schr\"odinger  type solutions
 describing nonrelativistic systems with varied dynamical exponents. 
The paper is organized as follows. In  section-II we introduce 
`boosted' black D$p$-brane solutions. These solutions have two 
parameters namely,
the horizon radius and the boost velocity, given by $r_0$ and $\gamma$. 
We  discuss a special  double limit
 in which the black hole horizon is allowed to shrink 
while the boost is simultaneously taken to be very large, but keeping 
their product fixed. The new  zero temperature geometries 
thus obtained describe  Lifshitz type dynamics with 
varied dynamical exponents as $a_{Lif}=2(p-6)/(p-5)$. 
In  section-III we study similar 
limits of the `boosted' bubble $p$-brane solutions. The 
resulting  solutions are instead of
 Schr\"odinger type solutions with 
dynamical exponents $a_{Sch}=2/(p-5)$. 
We do however find an interesting map 
between these Lifshitz and Schr\"odinger solutions under which 
dynamical exponents get mapped into each other. 
The section-IV we discuss thermodynamical aspects of Lifshitz
solutions and  singularities. 
We also report on an intriguing thermodynamical duality.
The conclusion is given in section-V. 

 \section{ Double limits of non-BPS D$p$-brane geometries }

We start with the black D$p$-brane  solutions  which are asymptotically
AdS solutions and  describe  temperature 
phases of the boundary CFTs. \footnote{ We shall be generally using the 
common phrases `AdS-black hole' and `AdS-Bubble' in this work, 
but the reader is alerted to
 keep in mind that the near horizon geometries such as \eqn{sol2} 
are usually of the type $\Omega^2(x)( ds^2[AdS_{p+2}] + ds^2[S^{8-p}])$ and 
there is a running dilaton field, except for $p=3$ case where conformal 
factor is constant. Nevertheless 
there is a well defined $(p+1)$-dimensional 
boundary CFT description on the boundary of $AdS_{p+2}$ subspace. We have 
also suitably rescaled $t, x^i$ in Eqs.(3) and (4) so as to absorve some  
$R_p$ factors.  } 
(We shall take $\alpha'=1$). These 
AdS black hole solutions are 
\bea\label{sol2}
&&ds^2_{BH}={ R_p^2 r^{p-3\over2 }}\bigg[ r^{5-p}(-f dt^2 +d\vec x_{(p)}^2)  
+{ dr^2\over f r^2}  + d\Omega_{(8-p)}^2 \bigg] ,\br
&& e^\phi=(2\pi)^{2-p}g_{YM}^2({R_p^4\over r^{7-p}})^{3-p\over4}\sim 
{\lambda_{eff}^{7-p\over 4}\over N }\ ,~~ ~~{\rm for} ~0\le p\le 6
\eea 
along with  electric flux for $(p+2)$-form 
field strength which couples to electrically charged $p$-branes $(p<3)$
as
\be
F_{p+2}\simeq (7-p) (R_p)^{2p-2} r^{6-p} dr\wedge dt\wedge [dx_{(p)}]\ .
\ee
For the magnetically charged solutions $(p>3)$ one instead  takes
 \be
F_{8-p}\simeq (7-p) (R_p)^4  \omega_{(8-p)},
\ee
 with $\omega_{(8-p)}$ 
being the volume form of a unit size rigid sphere, $S^{8-p}$. 
Especially for the D3-branes field strength $F_{(5)}$ should be taken
self-dual.
In the  metric $x_1,\cdots,x_p$ are flat spatial coordinates along
 the D$p$-brane 
world-volume, $N$ is the number of the branes, whereas
$d\Omega^2_{(8-p)}$ represents the 
line element of  unit  $(8-p)$-dimensional sphere, and 
 $(R_p)^2\equiv g_{YM}\sqrt{d_p N}$ represents radius square of the sphere. 
\footnote{ Here $d_p$ comprises suitable  combinatoric factor 
of 2's and $\pi$'s.}
In the function $f(r)=1-{r_0^{7-p}/r^{7-p}}$, $r=r_0$ is the size of
black hole horizon. 
Asymptotically as $r\to\infty$ the spacetime metric becomes conformal 
to  spacetime which is a direct
product of an $AdS_{p+2}$ and  $S^{8-p}$. 
The boundary conformal field theory will have  finite temperature 
where $t$ could be taken 
imaginary $(i\tau$) and suitably periodic, $\tau\sim \tau + {4\pi\over (7-p)}
r_0^{p-5\over 2}$. 
The effective 't Hooft coupling of the boundary SYM theory is given as
$$\lambda_{eff}\sim \lambda_0 r^{p-3},$$
  while $\lambda_0=g_{YM}^2 N$ being the bare 't Hooft coupling constant.
\footnote{
The details and early discussions on these AdS solutions can be found in 
\cite{itzhaki} and references therein.}
For our purpose here we have instead taken  noncompact Lorentzian time. 
We shall restrict to 
the branes with $1\le p\le 6$. For the D$0$-branes, as they are point like
we shall not study them here.

We make the following boost transformation involving one of the 
brane direction, say $y$,
\bea \label{boost54}
dt\to \gamma dt +v \gamma dy, ~~~
dy\to \gamma dy +v \gamma dt
\eea
where velocity $0\le v< 1~ (c=1)$ 
and the boost parameter $\gamma=1/\sqrt{1-v^2}$.
Since there is no Lorentz symmetry along all brane directions, particularly so
involving the $y$ direction, so we get a new
 boosted BH geometry 
\bea\label{sol2de}
&&ds^2=R_p^2{r^{p-3\over 2} }\bigg[ 
r^{5-p}\left(  
(1+{r_0^{7-p}v^2\g^2\over r^{7-p}})dy^2-
(1-{r_0^{7-p}\g^2\over r^{7-p}})dt^2+
{2 v\g^2 r_0^{7-p}\over r^{7-p}}dt dy
+d\vec x_{(p-1)}^2 \right)  \br
&&~~~~~~+{dr^2\over f r^2}  + d\Omega_{(8-p)}^2 \bigg] \br
&& e^\phi=(2\pi)^{2-p}g_{YM}^2({ r^{7-p}\over R_p^4})^{p-3\over4} \ , 
~~~{\rm for} ~1\le p\le 6
\eea 
while rest of the background fields remain unchanged under the boost, 
so we shall not writing them again here after, but it should be understood
that all branes are charged so we need appropriate flux. 
Here we would like to note that the periodicity of Euclidean $t$ in the 
boosted geometry gets modified and it is  now ${4\pi \g\over (7-p)}
 r_0^{p-5\over2}$. 
Correspondingly black hole horizon temperature is
\be
T= {(7-p) r_0^{5-p\over2}\over 4\pi \g}. 
\ee

\ud{\it The simultaneous $r_0\to 0,~\g\to\infty$ double limits:}

While setting $r_0=0$   in 
\eqn{sol2de} will give rise to usual 
AdS brane solutions, we  instead like to consider
a double limit in which $r_0$ is allowed to vanish
 while boost is simultaneously taken to be infinity, such that
\be\label{sol21}
r_0\to 0,~~~~ \g\to\infty, ~~~~r_0^{7-p}\g^2=\beta^2= {\rm fixed}.
\ee
 In which case we  have
\bea\label{sol22}
&&(1+f)\sim 2-O(r_0^{7-p}),~~~ {(1-f)\g^2}\to  {\beta^2\over r^{7-p}}
\eea
Introducing light-cone coordinates through $x^{\pm}=(t\pm y)$,
 the solution \eqn{sol2de} reduces to
\bea\label{sol2de1}
&&ds^2_{Lif}=R_p^2r^{p-3\over2}\bigg[ 
r^{5-p}( {\beta^2\over r^{7-p}} (dx^{-})^2 -dx^{+}dx^{-}
+d\vec x_{(p-1)}^2)  
+{dr^2\over  r^2}  + d\Omega_{(8-p)}^2 \bigg] \br
&& e^\phi=(2\pi)^{2-p}g_{YM}^2({ r^{7-p}\over R_p^4})^{p-3\over4},
\eea 
with suitable $$F_{p+2}=(7-p)R_p^{2p-2}r^{6-p} 
dr\wedge dx^+\wedge dx^-\wedge [dx_{(p-1)}]$$
for the electric type branes $(p<3)$ and 
$$F_{8-p}=(7-p) R_p^4 \omega_{8-p}$$ for magnetic type branes $(p>3)$.
 For D3-branes instead
we shall have $F_5=4 (1+\star)\omega_5$ which is self-dual.

The temperature vanishes as $r_0^{6-p}$ for $p<6$. 
Thus our limits imply vanishing temperature as $r_0\to0$ for $p<6$.
Especially for $p=6$, the temperature can be  arbitrary. 
We immediately notice that light-cone time is null, i.e. $g_{++}=0$,
 in  these  solutions.  
But a finite $g_{++}$ emerges once we compactify along $x^-$. 
(Since $g_{--}$ component is finite we can compactify along $x^-$ lightcone 
coordinate. This  would give rise to a Lifshitz like geometry in lower 
dimensions \cite{Singh:2010zs}. 
Also see \cite{donos10} for $a=2$ Lifshitz solutions where similar situations arise.) 
Nevertheless the  noncompact solutions \eqn{sol2de1}
 make complete solutions of type IIA/B string theory depending upon 
whether $p$ is even/odd. We can see that the 
$g_{--}$ component of the metric
is subdominant compared to $g_{+-}$ and $g_{ii}$ when 
$ r\to \infty$. 
Thus asymptotically the metric \eqn{sol2de} will become 
\bea\label{sol2d}
&&ds^2_{Lif}\sim R_p^2{r^{p-3\over2}}\left( r^{5-p}[  -dx^{+}dx^{-}
+d\vec x_{(p-1)}^2]  
+{dr^2\over  r^2}  + d\Omega_{(8-p)}^2 \right)
\br
&& e^\phi=(2\pi)^{2-p}g_{YM}^2({ r^{7-p}\over R_p^4})^{p-3\over4}
\eea 
along with a suitable $(p+2)$-form flux. 
These  precisely are the near horizon  (relativistic)
 geometries representing multiple $p$-brane systems. 
Thus we see that, for all $1\le p\le 6$, the
 nonrelativistic  deformations ($\beta$-terms in the metric) 
become prominent only in the
IR region. However this metric deformation will require us to switch on 
appropriate operators in the boundary field theory.
However, these operator deformations will disappear in the UV regime of 
corresponding CFT. Also due to these nonrelativistic deformations
in the IR regime, the scaling behaviours of 
space and time changes drastically. 
The time and space will now scale asymmetrically and we get 
Lifshitz like behaviour.

{\it Scaling symmetry:}

Let us redefine the radial coordinate as 
\be 
r^{p-5}=z^2.  
\ee
With $z$ as holographic  coordinate and some scalings the solutions 
can be written as (for $p\ne 5$)
\bea\label{sol2d03}
&&ds^2_{Lif}=R_p^2 z^{p-3\over p-5} \bigg[ 
( {\beta^2\over z^{4/(p-5)}} (dx^{-})^2 +{-dx^{+}dx^{-}
+d\vec x_{(p-1)}^2\over z^2}  
+({2\over 5-p})^2 {dz^2\over  z^2})  + d\Omega_{(8-p)}^2 \bigg] \br
&& e^\phi=(2\pi)^{2-p}g_{YM}^2R_p^{3-p}({ z^{2(7-p)\over p-5}})^{p-3\over4}
\eea 
with the $(p+2)$-form flux.
We find that under  asymmetric scalings (dilatation) of
the coordinates
\be\label{d1}
z\to \xi z , ~~~~
x^{-}\to \xi^{2-a}  x^{-},~~~~
x^{+}\to \xi^{a} x^{+},~~~\vec x\to \xi \vec x
\ee
with  dynamical exponent  $a={2(p-6)\over p-5}$,  the dilaton and 
the metric  in \eqn{sol2d03} conformally rescale as
\bea\label{d2}
g_{MN}\to \xi^{p-3\over p-5} g_{MN},
~~~e^\phi\to \xi^{(7-p)(p-3)\over2(p-5)}e^\phi
\eea
This is nothing but the standard  Weyl 
rescaling behaviour of  near horizon D$p$-branes AdS solutions \cite{itzhaki}. 
The Weyl scaling of the bulk metric and $e^\phi$ is indicative of the fact
that the boundary non-relativistic CFT (NRCFT) is not
a conformal theory instead it
 has got a running effective coupling $(\lambda_{eff})$.  
But what we find most surprising is the fact that the  RG flow 
is still of the standard type 
inspite of the asymmetric scalings of the coordinates. 
Only for  D3-branes  we  have a scaling symmetry 
which involves no Weyl rescaling of the bulk metric. 
For D3-brane case,  discussed earlier in \cite{Singh:2010zs},
one finds a Lifshitz spacetime
\bea\label{sol2d01}
&&ds^2_{D3}=R_3^2\bigg[ 
( {\beta^2\over z^{-2}} (dx^{-})^2 +{-dx^{+}dx^{-}
+d\vec x_{(2)}^2\over z^2})  
+{dz^2\over  z^2}  + d\Omega_{(5)}^2 \bigg] \br
&& e^\phi=(2\pi)^{-1}g_{YM}^2 \
, ~~~ F_{(5)}=4 R_3^4 (1+\star)\omega_{5}
\eea 
where the dynamical exponent  is $a=3$. 
These 5-dimensional spaces were previously also known as Kaigorodov 
spacetimes in literature or as `Einstein spaces of maximum mobility' 
\cite{kaig,clp}. 
The dual field theory thus involves an infinitely boosted CFT.
If we mainly
focus on the coordinate patch $(x^+,x^-,\vec x)$ at the boundary 
where the CFT lives,
we can see that the spacetime indeed possess  asymmetric scaling symmetry. 
 Since $x^-$ coordinate is not null,  
we can also think of compactifying  along this  direction.
In \cite{Singh:2010zs} 
it was shown that the lower dimensional solution resembles
a Lifshitz geometry.
To see this we could rewrite the metric of Eq.\eqn{sol2d01} in a 
diagonal basis as
\bea\label{sol2d02}
ds^2_{D3}
&=&R_3^2 \left( [-{(dx^{+})^2\over 4\beta^2z^6}
+{dx_1^2 +dx_2^2 +dz^2\over  z^2}]_{Lif_4} +
{\beta^2 z^2} 
(dx^{-}-{dx^{+}\over 2\beta^2 z^4})^2  + d\Omega_5^2\right) \br 
\eea 
From  metric Eq.\eqn{sol2d02} it is obvious that
 the nonrelativistic geometry indeed represents a  
system of  Kaluza-Klein particles (gravi-photons) 
in lower dimensions.
That is upon compactification, 
we will simply be dealing with, a dilaton and  graviphotons
in a 4-dimensional  Lifshitz universe 
$
\left( -{(dx^{+})^2\over 4\beta^2z^6}
+{dx_1^2 +dx_2^2 +dz^2\over  z^2}\right)$.
This story will repeat itself for all $p$-branes as and 
when we compactify the $x^-$ direction.

Note that once $x^{-}$ is compactified, 
$x^{-}\sim x^{-} + 2\pi r^{-}$, the Lifshitz 
geometry \eqn{sol2d01} acquires a complete distinct notion. They can
provide a valid holographic description but only in a definite $z$ range.
For example, solutions \eqn{sol2d02} cannot be trusted near the AdS boundary 
(in UV) because the physical size of  $x^{-}$ circle 
\be
{R^{-}_{phys}\over l_s}={R_3\over l_s} \beta{r^{-}z} 
\ee
becomes sub-stringy as $z\to 0$. 
This would necessitate higher derivative string (world-sheet) corrections 
to the solutions.  
Alternatively, as suggested in \cite{malda}, when such a situation arises, 
it will also be appropriate to go over to 
a T-dual type II string solution where the T-dualised $x^-$ circle 
can have a finite radius.

In addition to the scaling properties in Eqs.\eqn{d1},\eqn{d2} as 
described earlier, 
the Lifshitz solutions \eqn{sol2de1} or \eqn{sol2d03} also have
  invariances under the translations 
\be\label{inv1}
x^+ \to x^+ + b^{+},~~~~
x^- \to x^- + b^{-},~~~~x^i\to x^i +b^i
\ee
 and  under the rotations of $x^i$  coordinates. 
 However, due to nontrivial $g_{--}$ components these
 spaces \eqn{sol2d01} do not  have  any explicit invariance under
the  Galilean boost 
\be\label{boos}
x^+\to x^+,~~~x^-\to x^- -2\vec v.\vec x +{v^2}x^+,~~~\vec 
x\to~\vec x-\vec v x^+ \ ,
\ee
$\vec v$ is constant velocity.
However there exists an unusual symmetry under space-like shifts
of the time $(x^+)$
\be\label{boos6}
x^-\to x^-,~~~x^+\to x^+ -2\vec v.\vec x +{v^2}x^-,~~~\vec 
x\to~\vec x-\vec v x^-.
\ee
The latter shift symmetry will be absent when $x^-$ coordinate  is compactified.
Thus
 our noncompact solutions \eqn{sol2d03} represent Lifshitz geometry 
when compactified along $x^-$, in which the time scales asymmetrically with 
dynamical exponent  
\be\label{d3}
a_{Lif}=2{p-6 \over p-5}.
\ee
Note, however, no extra matter fields are present in
the backgrounds \eqn{sol2de1} except dilaton and  $(p+2)$-form fluxes. 

Let us now find the supersymmetries preserved by the Lifshitz solutions 
\eqn{sol2de1}. Since all these solutions are in the same class, 
let us pick up the simplest case of D$3$-branes. 
We have checked that for $p=3$  Lifshitz backgrounds \eqn{sol2de1}
 all  superconformal Killing spinors identically vanish, while remaining
sixteen Poincare\'{}  Killing spinors have to satisfy an additional  
condition $\Gamma^+ \epsilon=0$. 
Thus only 8  Poincare\'{}  supersymmetries survive for the Lifshitz background, 
so they are only $1/4$-BPS solutions, see also \cite{Singh:2010zs}. 
The same would be true for the case of $p$ other than 3.

Separately, for D5-branes we get the solution  (following from \eqn{sol2de1}
and defining $r=1/z$) as
\bea\label{sol2d5}
&&ds^2_{D5}=R_5^2 {1\over z} \bigg[ 
( {\beta^2 z^{2}} (dx^{-})^2 -dx^{+}dx^{-}
+d\vec x_{(4)}^2)  
+{dz^2\over  z^2}  + d\Omega_{(3)}^2 \bigg] \br
&& e^\phi=(2\pi)^{-3}{g_{YM}^2 \over R_5^2 z}
,~~~ F_{(3)}=2 R_5^4 \omega_{(3)}
\eea 
where Lifshitz scaling is $z\to \xi z,~x^+\to\xi x^+,~x^-\to \xi^{-1}
x^-, ~x^i\to x^i$ along with the Weyl rescalings. Note $x^i$'s do not need to 
rescale in these 5-brane solutions and 
they have dynamical exponent $a=1$. So they appear to
remain relativistic branes, but they are not due to
 nontrivial $g_{--}$ metric component. 

For D6-branes from \eqn{sol2d03} we have
\bea\label{sol2d03hh}
&&ds^2_{D6}=R_6^2 z^{3} \bigg[ 
( {\beta^2\over z^{4}} (dx^{-})^2 +{-dx^{+}dx^{-}
+d\vec x_{(5)}^2\over z^2}  
+4{dz^2\over  z^2})  + d\Omega_{(2)}^2 \bigg] \br
&& e^\phi=(2\pi)^{-4}g_{YM}^2({ z\over R_6^2})^{3\over2}
\eea 
where scaling invariance is $z\to \xi z,~x^+\to x^+,~x^-\to \xi^{2}
x^-, ~x^i\to \xi x^i$ along with Weyl rescaling. Note $x^+$ does not need to 
rescale in these 6-brane Lifshitz solutions, so 
they have dynamical exponent of time as $a=0$. So for these solutions,
 their Euclidean counter part can have $x^+$ with arbitrary periodicity,
which means an arbitrary unfixed temperature.

\section{ Double limits of AdS-Bubble  geometries and 
Schr\"odinger spacetime}

Here we start with the `bubble' spacetimes  which are asymptotically
$AdS$ solutions of type II string theory.
 The AdS-bubble solutions are well known to describe the low temperature 
phases in their respective holographic SYM theories. 
These bubble geometries are 
\bea\label{sol3h}
&&ds^2_{Bubble}=R_p^2 {r^{p-3\over2 }}\bigg[ r^{5-p}(-dt^2+fdy^2 +
d\vec x_{(p-1)}^2)  
+{ dr^2\over f r^2}  + d\Omega_{(8-p)}^2 \bigg] ,\br
&& e^\phi=(2\pi)^{2-p}g_{YM}^2({r^{7-p}\over R_p^4})^{p-3\over4}
\eea 
with appropriate  flux for $F_{p+2}$ form  field strength.
We shall consider again  $1\le p\le 6$. 
We note here that the above bubble p-brane geometries can be obtained 
by performing a double Wick rotation  
\be\label{wick1}
t\to iy, ~~~~y\to i t
\ee
of the black hole solutions given in \eqn{sol2}. But they make two physically 
distinct solutions, one involves black holes and the others do not.  
In the above $f(r)=(1-{r_b^{7-p}/r^{7-p}})$ with $r$ range being
$r_b\le r\le \infty$. Since the radial coordinate is restricted in the IR
region, they describe what is commonly known as `bubble geometry', 
inside of the bubble is empty. 
The coordinate $y$ has to have correct periodicity, 
$y\sim y+ {4\pi\over (7-p)}r_b^{p-5\over2}$, 
in order  that metric avoids the singularity 
at $r=r_b$ and assumes a  sigar 
like spatial geometry.  
Asymptotically bubble spacetimes become 
 conformal to a geometry which is a product of an  AdS spacetime with 
a compact spatial coordinate and a round sphere, and 
having a running dilaton field, 
 except for the case of $p=3$ case where dilaton becomes constant. 
The boundary is at $r\to\infty$. 
Since $r$ is the holographic (energy) coordinate,
the boundary conformal field theory has an effective IR cut-off scale as
$r=r_b$. The field theory can  have 
finite small temperature if the Euclidean $t$ is taken 
to be  periodic. The period of Euclidean time can be 
arbitrary in these solutions but it 
is usually fixed by the temperature of the boundary CFT.
  Here we have taken Lorentzian time  for the purpose of this work. 

We employ the following boost transformation along compact $y$ coordinate
\bea 
dt\to \gamma dt + v \gamma dy, ~~~
dy\to \gamma dy + v \gamma dt
\eea
with boost parameter $\gamma=1/\sqrt{1-v^2}$.
Since there is no Lorentz symmetry involving $y$ direction, we get a
 boosted bubble metric
\bea\label{sol3h1}
&&ds^2=R_p^2{r^{p-3\over2}}\bigg[ 
r^{5-p}\left( - 
(1+{r_b^{7-p}v^2\g^2\over r^{7-p}})dt^2+
(1-{r_b^{7-p}\g^2\over r^{7-p}})dy^2+
{2 v\g^2 r_b^{7-p}\over r^{7-p}}dt dy
+d\vec x_{(p-1)}^2 \right)  \br
&&+{dr^2\over f r^2}  + d\Omega_{(8-p)}^2 \bigg]
\eea 
while other background fields remain unchanged under the boost. 
Here we would like to note that the  $y$-periodicity in the 
boosted geometry gets modified to 
$y\sim y+ {4\pi\gamma\over (7-p)}r_b^{p-5\over2}$, 

\ud{\it The simultaneous $r_b\to 0,~\g\to\infty$  limit:}

In order to find nonrelativistic brane solutions
we  consider
 double limits in which $r_b$ is allowed to vanish
while boost is simultaneously taken to be large, such that
\be\label{sol3h2}
r_b\to 0,~~~~ \g\to\infty, ~~~~r_b^{7-p}\g^2=\beta^2= {\rm fixed}.
\ee
Note that while we take these double  limits 
$y$ coordinate indeed gets decompactified $(p<6)$. In lightcone coordinates 
 the solution \eqn{sol3h1} reduces to
\bea\label{sol3h3}
&&ds^2_{Sch}=R_p^2 r^{p-3\over2} \bigg[ 
r^{5-p}( -{\beta^2\over r^{7-p}} (dx^{+})^2 -dx^{+}dx^{-}
+d\vec x_{(p-1)}^2)  
+{dr^2\over  r^2}  + d\Omega_{(8-p)}^2 \bigg] \br
&& e^\phi=(2\pi)^{2-p}g_{YM}^2({R_p^4\over r^{7-p}})^{3-p\over4}
\eea 
with appropriate $(p+2)$-form flux. 
We can notice that $x^-$ is null in these Schr\"odinger type solutions. Which 
is juxtapose of the Lifshitz like solutions  obtained earlier from AdS black
holes. Although, $x^-$ is an isometry direction, but
as it remains a null direction, 
it will not be possible to compactify along this lightcone 
coordinate. Indeed the situation is much like  for Schr\"odinger solutions 
with dynamical exponent $2$, see  Maldacena et.al \cite{malda}.
Nevertheless the Schr\"odinger
solutions \eqn{sol3h3} are  complete solutions of type II string 
theory. Asymptotically
 in the region where $ r^{7-p} \gg \beta^2$ the metrics 
precisely become  AdS geometries representing
 multiple $p$-branes. 
Thus  for all $p\le 6$ the
 nonrelativistic (Schroedinger type) deformations (namely $g_{++}$ components)
become prominent only in the
IR region. While all  NR effects disappear in the UV regime. 

{\it Asymmetric scaling symmetry:}

Let us introduce the holographic $z$-coordinate again. With $z$ coordinate 
the solutions become (for $p\ne 5$)
\bea\label{sol3h4}
&&ds^2_{Sch}=R_p^2 z^{p-3\over p-5} \bigg[ 
[ -{\beta^2\over z^{4\over p-5}} (dx^{+})^2 +{-dx^{+}dx^{-}
+d\vec x_{(p-1)}^2\over z^2}  
+{4\over (5-p)^2}{dz^2\over  z^2}]  + d\Omega_{(8-p)}^2 \bigg] \br
&& e^\phi=(2\pi)^{2-p}g_{YM}^2({R_p^4 z^{2(p-7)\over p-5}})^{3-p\over4}
\eea 
with  $(p+2)$ form flux.
Thus we see that there is an asymmetric scaling invariance involving 
the coordinates
\be
z\to \xi z , ~~~~
x^{-}\to \xi^{2-a}  x^{-},~~~~
x^{+}\to \xi^{a} x^{+},~~~\vec x_{p-1}\to \xi \vec x_{p-1}
\ee
with  dynamical exponent 
\be
a_{Sch}={2\over p-5},
\ee where dilaton and the  metric conformally
scale as
\bea
g_{MN}\to \xi^{p-3\over p-5} g_{MN},
~~~e^\phi\to \xi^{(7-p)(p-3)\over2(p-5)}e^\phi
\eea
It is nothing but precisely the usual Weyl scaling
 behaviour of the Dp-branes near horizon AdS geometries. 
If we focus only on the coordinate patch 
$(x^+,x^-,\vec x)$ where the CFT lives,
 we see that the 
 CFT will  exhibit Schr\"odinger symmetry  
but devoid of special conformal symmetry. 
The solutions \eqn{sol3h4} are also invariant 
under translations, rotations and Galilean boosts 
Eqs. \eqn{inv1} and \eqn{boos}.

Especially for D3-branes case we find
\bea\label{sol2dd3}
&&ds^2_{D3}= R_3^2 \bigg[ 
 -{\beta^2 z^2} (dx^{+})^2 +{-dx^{+}dx^{-}
+d\vec x_{(2)}^2\over z^2}
+{dz^2\over  z^2}  + d\Omega_{(5)}^2 \bigg] \br
&& e^\phi=(2\pi)^{-1}g_{YM}^2
,~ ~~ F_{(5)}=4 R_3^4(1+\star)\omega_5
\eea 
which is a Schr\"odinger  spacetime but with the dynamical 
exponent as $a=-1$.
Let us note that Schr\"odinger spacetimes with dynamical exponent  $a=2$  
only admit special conformal symmetry \cite{son,bala}. 

The Schr\"odinger spacetime with $a=-1$ can 
also be called a Kaigorodov space. 
The Kaigorodov spaces are known as Einstein
spacetimes of maximum mobility \cite{kaig}. Corresponding
boundary dual theories have been described as CFTs in infinite 
momentum frame \cite{clp}.
So it is not surprising that we obtained our solutions from boosted
`black' and `bubble' $p$-brane solutions
under the limits involving infinite  boosts. For example,
the  Ricci tensor for the 5D Kaigorodov space is
simply given by $R_{\mu\nu}= -4 g_{\mu\nu}$, and so 
the curvature scalar is
 just a constant. Given only these quantities the Schr\"odinger spacetimes 
 would look just like an ordinary $AdS_5$ space, however the
Weyl tensor for Schr\"odinger-Kaigorodov spaces remains nonvanishing, 
see also Appendix. 
Thus unlike pure AdS spaces which are conformally flat, 
the Schr\"odinger spaces will not be so. 
Also the isometries and local structure of spacetimes
are completely different, see \cite{clp}.  

Especially for D6-branes from \eqn{sol3h4} we have
\bea\label{sol3h4hh}
&&ds^2_{D6}=R_6^2 z^{3} \bigg[ 
( -{\beta^2\over z^{4}} (dx^{+})^2 +{-dx^{+}dx^{-}
+d\vec x_{(5)}^2\over z^2})  
+4{dz^2\over  z^2}  + d\Omega_{(2)}^2 \bigg] \br
&& e^\phi=(2\pi)^{-4}g_{YM}^2({ z\over R_6^2})^{3\over2}
\eea 
where scaling invariance involves $z\to \xi z,~x^-\to x^-,~x^+\to \xi^{2}
x^+, ~x^i\to \xi x^i$. Note $x^-$ does not need to 
rescale in these 6-brane Lifshitz solutions, so 
they  have a dynamical exponent $a=2$. So only for D6 solutions,
 we can have $x^-$ with arbitrary periodicity. Although these solutions have
interesting value $a=2$, 
but there is an overall Weyl scaling of metric and dilaton required.

\subsection{A relationship of dynamical exponents}

An intriguing but interesting aspect of our Lifshitz and Schr\"odinger 
$p$-brane solutions, in eqs. \eqn{sol2d03} and \eqn{sol3h4}, is that
they can be mapped into each other under the following
exchange of the light-cone metric components
\be\label{exchan1}
g_{++}^{(Lif)}\to -g_{--}^{(Sch)},~~~~
g_{--}^{(Lif)}\to -g_{++}^{(Sch)} . 
\ee
While  respective dynamical exponents get  mapped as
\be\label{gy7}
a_{Lif}=2-a_{Sch}
\ee
for all the D$p$-branes. We have $a_{Lif}=2({p-6\over p-5})$ while
$a_{Sch}={2\over p-5}$, for all $0<p\le 6$ but $p\ne5$. Especially, 
the D5-branes have $a_{Lif}=1$. 
So the relation \eqn{gy7} is obeyed by all nonrelativistic
 D$p$-branes presented  here. This could be understood as follows. 
Under the Wick rotations \eqn{wick1}, which takes AdS-BH into 
AdS-Bubble and vice versa, 
the lightcone coordinates map as
\be\label{wick2}
x^+\to -i x^-,~~~
x^-\to i x^+.
\ee
The above double Wick  rotation \footnote{ We are thankful to 
the referee to have suggested this.}
will exactly perform the operation 
given in \eqn{exchan1},  
keeping $g_{+-}$ fixed. 
 Although related mathematically in this simple manner, 
the Lifshitz and Schr\"odinger $p$-Brane solutions represent physically 
distinct spacetimes, having different dynamical exponents. Thus it is 
not surprising that just as AdS-BH and Bubble solutions get mapped into 
eath other under double Wick rotations, the Lifshitz \eqn{sol2d03}
and Schr\"odinger \eqn{sol3h4}
spacetimes do as well map into each other.  
A comparative list of dynamical exponents is provided  
in the table \eqn{Table1}.

\begin{table}[h]
\begin{center} 
\begin{tabular}{cccc}
\hline $p$-brane & $a_{Lif}$& $ a_{Sch}$& $(a_{Lif}+a_{Sch})$ \\ \hline 
 \hline
1 & 5/2 & - 1/2 &2 \\ \hline
2 & 8/3 & -2/3 &2 \\ \hline
3 & 3 & -1 &2 \\ \hline
4 & 4 & -2 &2 \\ \hline
5 & 1 & 1 &2 \\ \hline
6 & 0 & 2 &2 \\ \hline
\end{tabular} 
\caption{\label{Table1} Dynamical scaling exponents of the 
Lifshitz and the Schr\"odinger
solutions}
\end{center}
\end{table}
It is interesting to notice that the dynamical exponents of Lifshitz 
geometries are all positive. Especially, for 3-branes it is $a=3$. We  
study it a bit further. Compactifying the Lifshitz solution
\eqn{sol2d02} along $x^-$ and $S^5$, the remaining 4-dimensional Lifshitz 
spacetime can be written in the Einstein frame as
\be\label{pok9}
ds^2_{Lif_4}\sim z [-{(dx^{+})^2\over \beta^2z^6}
+{dx_1^2 +dx_2^2 +dz^2\over  z^2}].
\ee
Since the number of spatial dimensions is $d=2$ here, the Lifshitz geometry
\eqn{pok9} indeed represents
 a spacetime which has a hyperscaling  exponent given as $\theta=1$, 
and under the asymmetric scalings $
x^+\to \lambda^3 x^+,~
x^1\to \lambda x^1,~
x^2\to \lambda x^2,~
r\to \lambda r$
the metric \eqn{pok9} scales as 
$$ds^2\to \lambda^{2\theta\over d} ds^2 .$$ 
We also note that this $a=3$ solution satisfies an important criterion 
 $a\ge {\theta\over d}+1$ and also has
 $\theta=d-1$. These conditions 
involve putting Null energy conditions on the energy-momentum tensor
 in the Lifshitz spacetime. 
These studies have been a subject of much attention recently in 
the literature, see  
\cite{takaya11,subir11}.

Similarly for other Lifshitz $p$-brane solutions in \eqn{sol2d03}, 
after compactifications along $x^{-}$ and $S^{8-p}$,
 we find that the hyperscaling dimensions are
\be
\theta_{(p)}={p^2-6p+7\over p-5}, ~~~~p\ne5.
\ee
and corresponding $(p+1)$-dimensional Lifshitz geometries can be written 
in Einstein frame as
\be\label{pok9a}
ds^2_{Lif_{p+1}}\sim
z^{2(p^2-6p+7)\over (p-1)(p-5)}\left( -{(dx^{+})^2\over \beta^2 
z^{4(p-6)\over (p-5)}}
+{(d\vec{x}_{p-1})^2  +dz^2\over  z^2}\right).
\ee
We find that physically interesting cases of $p=2,3,4$ they all satisfy 
$a\ge {\theta\over d}+1$. These correspond to the cases where boundary 
CFTs have spatial dimensions $d=1,2,3$ respectively. 
The Schr\"odinger geometries in the table are rather puzzling as all lower
dimensional cases have negative dynamical exponents,  except 
for $p=5$ and $p=6$. Also the spatial lightcone coordinate $x^-$ 
is null there, so we 
cannot compactify along this direction. The problem we face here
is similar to the case of Schr\"odinger solutions in \cite{malda}.
It would be worth while to explore them further and we hope to 
return to them in near future.

\section{ Lifshitz singularities and thermodynamics }

It has been recently shown  that Lifshitz  solutions have 
essential null curvature singularities \cite{Horowitz:2011gh}. It is
due to fact that a test string become infinitely excited 
as it nears $r=0$, even though the  curvature  scalar 
 may be uniformally constant for the Lifshitz solutions (at least
for the case of D3-branes). These instabilities would necessitate us to 
include higher (derivative) order  
corrections to these classical  backgrounds. 
However, it would also be a safer
approach to hide these essential Lifshitz singularities 
behind some event horizon. Correspondingly, the finite temperature 
Lifshitz solutions will be those which have finite horizon size 
$r_0$ and also finite boost. Let us write them down explicitely
following from \eqn{sol2de} (direction of the boost is taken 
to be along negative $y$ direction)
\bea\label{sol2dq1}
ds^2&=&R_p^2{r^{p-3\over 2} }\bigg[ 
r^{5-p}\left(  
(1+{r_0^{7-p}v^2\g^2\over r^{7-p}})dy^2-
(1-{r_0^{7-p}\g^2\over r^{7-p}})dt^2-
{2 v\g^2 r_0^{7-p}\over r^{7-p}}dt dy
+d\vec x_{(p-1)}^2 \right)  \br
&&~~~~~~+{dr^2\over f r^2}  + d\Omega_{(8-p)}^2 \bigg] \br
 e^\phi&=&(2\pi)^{2-p}g_{YM}^2({ r^{7-p}\over R_p^4})^{p-3\over4} \ , 
\eea 
along with  $(p+2)$-form flux. 
Introducing lightcone coordinates, $x^{\pm}=t\pm y$,
the metric can be reexpressed  as
\bea\label{sol3a1}
ds^2&=&
R_p^2 r^{p-3\over2} \bigg[
r^{5-p}\big\{ ({r_0^{7-p}\over 4 r^{7-p}} {(dx^{+})^2\over\lambda^2}+
{r_0^{7-p}\lambda^2\over 4 r^{7-p}} (dx^{-})^2-{1+f\over2}dx^+dx^-)
+d\vec x_{(p-1)}^2\big\}\br &&~~~~~+{dr^2\over f r^2}  + d\Omega_{(8-p)}^2\bigg]
\br
&=&R_p^2 r^{p-3\over2} \bigg[ \big\{
r^{5-p}( -f {r^{7-p}\over W^2} (dx^{+})^2 
+d\vec x_{(p-1)}^2)+{dr^2\over f r^2}\big\}  + d\Omega_{(8-p)}^2
\br &&~~~~~~ +{W^2\over 4r^2}(dx^- -{1+f\over  r^{p-7}}{dx^+ \over W^2})^2 
 \bigg] .
\eea 
The parameters are related as
 $W^2\equiv r_0^{7-p}\lambda^2$
and the function $$f=(1-{r_0^{7-p}\over r^{7-p}}),$$  
and $r=r_0$ is the horizon.\footnote{ The boost parameter $\lambda \ge 1$ 
is written as $\lambda= \sqrt{1+v \over 1-v}$.} \footnote{ To clarify,
we have purposely  written down boosted black brane geometries  
in the form as in \eqn{sol3a1} because we
are interested in studying these solutions when $x^-$ is compact 
(coordinate on a circle). Clearly,
the remaining noncompact part of the metric resembles the
thermal Lifshitz solutions. Note
the thermal expressions in \eqn{tq} are for such a  3-brane
solution, with $x^-$ being compact.}
Note that we can rescale  $W$  out of 
these BH solutions, which is possible if we exploit the lightcone scaling
$x^+\to W x^+,~x^-\to W^{-1}x^-$. So we can set $W$ to unity if we wish 
but it is an useful parameter in the following analysis.  
For very large $W$ value these  black holes  in the intermediate range 
$r_0<r < W^{2\over{7-p}} $ 
will always behave as thermal Lifshitz like solutions. Let us  call this region
as the 
Lifshitz window region
where parameter $W$ provides the effective width of this window. 
 While
in the deep UV region, $r\gg W^{2\over{7-p}} $ the solutions 
become asymptotically AdS, see the figure \eqn{figure1}.
 \begin{figure}[h]
\centerline{\epsfxsize=5in
\epsffile{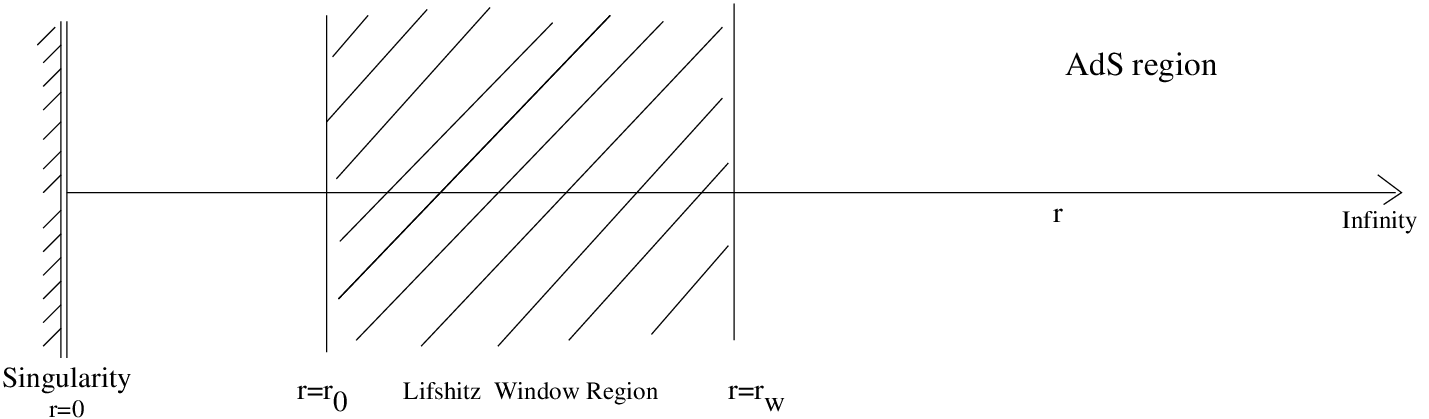} }
\caption{\label{figure1} The Lifshitz  
window appears as the shaded region. It starts at $r_0$ and ends at
$r_w \sim W^{2\over{7-p}}$.}
\end{figure}
\begin{figure}[h]
\centerline{\epsfxsize=5in
\epsffile{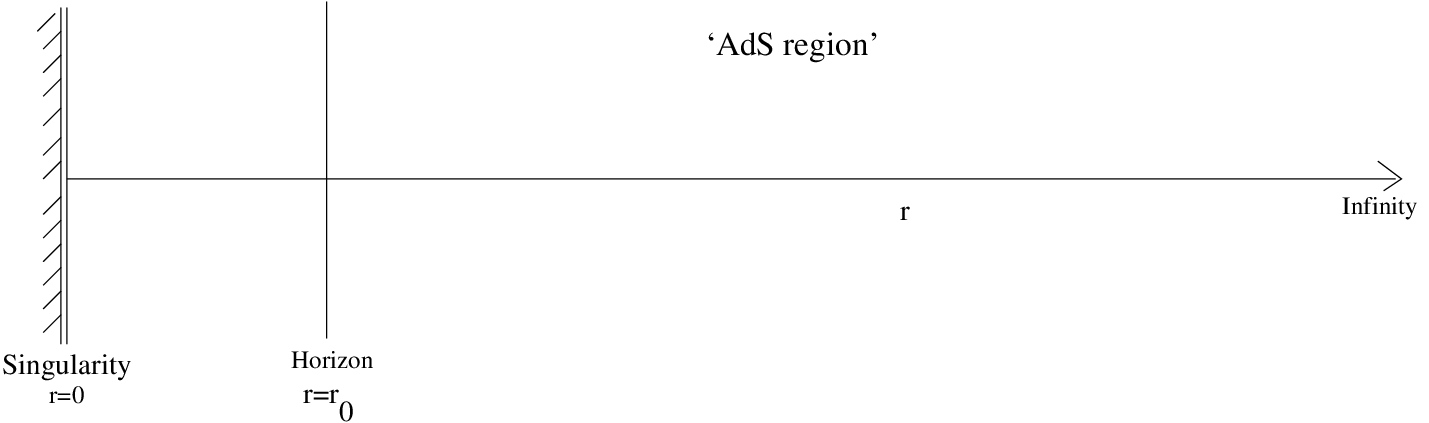} }
\caption{\label{figure2} Depiction of an usual AdS black hole spacetime. 
There is no Lifshitz like region here.}
\end{figure}
\begin{figure}[h]
\centerline{\epsfxsize=5in
\epsffile{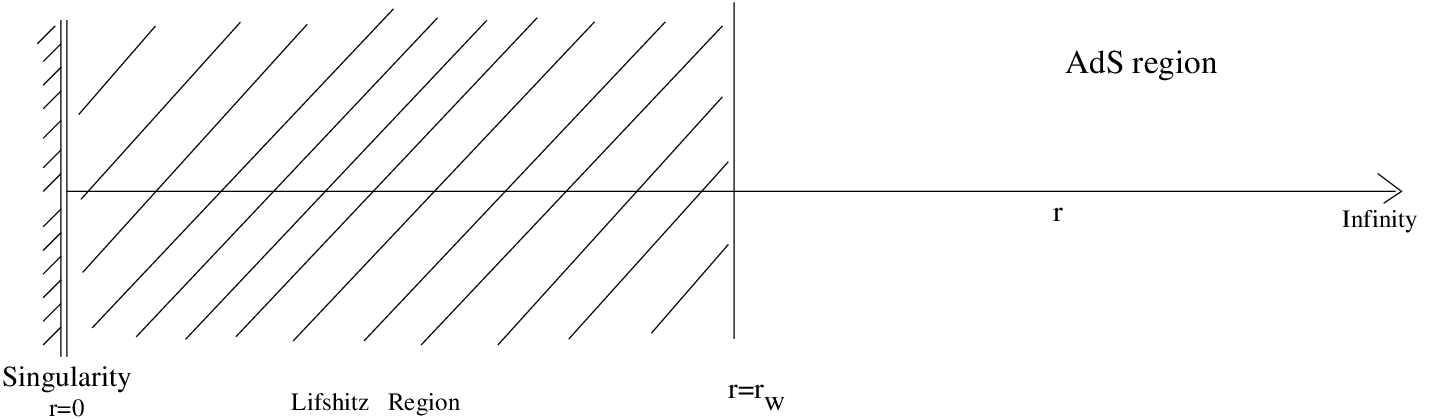} }
\caption{\label{figure3} In zero temperature solutions the Lifshitz  
window (the shaded region) starts at $r=0$ and ends at
$r_w$.}
\end{figure}
Note that the size of Lifshitz window  can be widened if we take 
$\lambda$ (boost) sufficiently large. But if $\lambda=1$ 
the Lifshitz region altogether disappears and we get ordinary 
AdS-BH solutions; see figure \eqn{figure2}.
We mentioned it earlier also that all our Lifshitz 
solutions become  AdS spaces asymptotically.  
These  Lifshitz BH solutions \eqn{sol3a1} with an 
intermediate Lifshitz region should
present a good IR description (at finite temperature), 
where the black hole horizon size provides an effective
 IR (thermal) cut-off in the dual CFT. Of course, there is a black hole 
singularity but it will be hidden behind a horizon and cannot be seen by 
an asymptotic observer. So the issue of curvature singularity appears to be fine
for black hole  Lifshitz solutions \eqn{sol3a1}. On the other hand
our double limits  in Eq.\eqn{sol21},
however correspond to the widening of the Lifshitz window such that the 
left edge of the window moves to $r=0$ while the right edge is held fixed 
at $r=r_w$. Consequently the limits give rise to zero 
temperature Lifshitz solutions 
given in equations \eqn{sol2de} and \eqn{sol2de1}. 
The zero temperature Lifshitz region is sketched in figure \eqn{figure3}.

\subsection{Double limits and  thermodynamic duality}

Having obtained the  nonrelativistic geometries
We shall need to study the effect of our double limits on the thermodynamic 
quantities. These finite temperature boundary field theories involve DLCQ 
description  \cite{malda}. For our Lifshitz spacetimes,
the compactification along $x^{-}$  implies that there is a conserved
 charge (momentum) $P_{-}$ which is quantized in units of 
${1\over r^-}$. The number (momentum) density  depends upon 
the choice of two parameters, namely boost $\gamma$ (or $\lambda$) and $r_0$.   
It would be worth while to  know what happens to the
 number density, energy density
$(-P_{+})$ and other thermodynamical quantities; like temperature $(T)$, 
entropy $(S)$ 
and chemical potential $(\mu_{_N})$, relevant
for the black-hole solutions, as we consider the double  limits. 
 These thermodynamic expressions are  considered here for 
the case of D3-branes for concreteness
\bea\label{tq}
&& \rho={N\over V_2}={r^{-}(-P_{-})\over v_2}
={L^3\over G_N^5}{(r^{-})^2\lambda^2 
r_0^4\over 8}     \br
&& {\cal E}={H\over V_2}={(-P_{+})\over v_2}={L^3\over G_N^5}{r^{-} 
r_0^4\over 16}\br
&&s= {S\over V_2}={L^3\over4 G_N^5}(2\pi r^{-}){\lambda 
r_0^3\over 2}  \br
&& { T}={r_0\over \pi \lambda},
~~~~~\mu_{_N}={1\over r^{-}\lambda^2} 
  \eea
where $L$ represents the $AdS_5$ radius, 
$G_N^5$ is 5-dimensional Newton's constant
and $V_2$ is some finite volume of 
$x_1-x_2$ plane. These quantities eventually  
satisfy the first law of thermodynamics 
\be
\delta E(T,\mu_{_N})=T \delta S-\mu_{_N}\delta N
\ee
We see that under the 
  limits \eqn{sol21}, 
the temperature of  boundary $(2+1)$ dimensional 
nonrelativistic field theory effectively  vanishes,
so also the entropy and the chemical potential.
Although  these quantities are vanishing,  
worth noticing is their unique scaling behaviour as powers of 
vanishing horizon size $r_0$, observed initially in  \cite{Singh:2010zs},
\be\label{tq1}
T\sim r_0^3\sim 0, ~~~{\cal \rho}= {\rm fixed}, ~~~~\mu_{_N}\sim r_0^4\sim 
0, ~~~~s\sim r_0 \sim 0.
\ee
Especially,  the temperature vanishes as cubic power of 
$r_0$,
which is an indication of the fact that system behaves nonrelativistically
 having
 dynamical exponent $a=3$ as it undergoes a condensation in the  DLCQ  theory. 
\cite{Singh:2010zs}.
We can however reexpress \eqn{tq1} as  
 \be
T\to 0,~~\mu_{_N}\to T^{4\over 3}\sim 0,~~s\to T^{1\over 3}\sim0,~~\rho \sim {T^4\over  \mu_{_N}^3}={\rm fixed}
\ee
These kind of strange behaviours, $s\sim T^{1\over 3}$, and specific heat
$C= T{\partial S\over \partial T}\sim T^{1\over 3}$,
have  recently been related to some non-Fermi liquids (strage metals) 
 in 2d systems \cite{takaya11,subir11}.
Curiously  though, we  observe that  the  number 
density has to remain fixed in order to achieve this condensation. Obviously
the energy density is also  vanishing. 
We  may also write down  
expressions for free energy density and the entropy density 
\bea
&& F\sim -{\cal E}\sim -{T^4\over\mu_{_N}^2}\sim r_0^4\sim 0,~~~~
s \sim {T^3\over  \mu_{_N}^2}\sim r_0\sim 0
\eea 

Let us compare above double limits i) \underline{$r_0\to 0,~\lambda=\infty$ } 
with 
the ii) \underline{$r_0\to 0,~\lambda=$ finite}   limit. 
Under the latter kind of
vanishing horizon (extremal) limit, the thermodynamical expressions behave as
\be\label{tq1e}
T\sim r_0\sim 0, ~~~
 \mu_{_N}\sim {\rm fixed},~~~{\rho}\sim  r_0^4 \sim 0  
~~~s\sim r_0^3 \sim 0
\ee
and ${\cal E}\sim r_0^4\sim 0$, as horizon size shrinks to zero.
Particularly,  the temperature vanishes as unit power of $r_0$
which is an indication of the fact that system behaves ordinarily 
relativistic, having 
 dynamical exponent $a=1$,  as it undergoes a  condensation.
Juxtapose to the double limit case of \eqn{tq1},
  we  find here that  the chemical potential in \eqn{tq1e}
remains fixed while the number (momentum) density  becomes 
vanishing, $O(r_0^4)$, in the relativistic situation. 
 
From the above rather crude exercise and
from the power law behaviour of the thermal quantities in \eqn{tq1} 
and \eqn{tq1e},  
we would infer that two different limits describe
 two distinct types of condensation phenomenon; non-relativistic 
(Lifshitz) and the 
relativistic (Lorentzian). However, these two condensation  points 
appear to be thermodynamically {\it dual} points. They
 could be Legendre transformed into each other, 
\bea
T\leftrightarrow S,~~~~ 
\mu_{_N} 
\leftrightarrow N, ~~~{\cal E}(T,\mu_{_N}) \leftrightarrow \tilde{\cal E}(S,N).
 \eea
Since Legendre transformations of thermodynamic variables 
do map  micro-canonical ensemble
into a canonical/grand-canonical ensemble and vice versa, 
the two types of condensation ($T\to 0$) limits described above seem to 
represent two different physical ensembles near respective fixed points. 
We indeed observe that the condensation behaviour of the
 canonically conjugate  variables, $(T,S)$ and $(\mu_{_N},\rho)$ 
does seem to get mapped into each other. 

\section{Conclusion}

We have extended our previous study of D3-branes  where vanishing horizon
double limits of 
`boosted' $AdS_5\times S^5$ black holes gave rise to Lifshitz type solutions
 with dynamical exponent $a=3$
\cite{Singh:2010zs}.  Taking the similar limits for all 
D$p$-branes,
we obtain  a wide class of Lifshitz like solutions. However the asymmetric
scaling of the coordinates has to be accompanied by a Weyl rescaling of the   
metric and dilaton field etc. This rescaling is identical to the standard  Weyl 
scaling behaviour of  D$p$-branes AdS metrics \cite{itzhaki}. Thus 
the Weyl rescaling of the string metric and $e^\phi$ is indicative of the fact
that the boundary NRCFT  is not  conformal instead it
 has got  running effective couplings.  
But what we find most interesting is that this  RG flow of the NRCFT
is still of the standard type. 
The new geometries 
so obtained describe nonrelativistic Lifshitz type dynamics with 
varied dynamical exponents, $a_{Lif}=2{p-6\over p-5}$. 
We have also studied  similar double
limits for  boosted `bubble' D$p$-brane solutions. The 
resulting Galilean solutions instead are classified as
 Schr\"odinger type, with 
varied dynamical exponents as $a_{Sch}={2\over p-5}$. 
We did find an interesting map 
between these Lifshitz and Schr\"odinger type solutions, 
such that for a given $p$-brane case
 respective dynamical exponents get related as
$$a_{Lif}=2-a_{Sch}\ .$$
 
We also discussed  thermodynamical aspects involving Lifshitz
black hole solutions and the issue of null curvature singularities. 
We find that in order to study Lifshitz systems at finite temperature
one can include black holes. In this way we can avoid seeing the essential 
Lifshitz singularities.    
Indepedently an observation regarding  an intriguing thermodynamical duality
is also reported.
  
As we were typesetting this work a few papers \cite{naray,kim} appeared 
in last couple of days which might see some overlap with us. 

\vskip.5cm
\noindent{\it Acknowledgements:}

I take this opportunity to thank the organisers of 
the `National Strings Meeting 2011',
University of Delhi, for the warm hospitality where part of this work
was carried out.

\appendix{

\section{Double limits of some flat space solutions and plane waves}

Consider a Schwarzschild black hole solution in $D$ spacetime dimensions 
which is delocalised (isometry) along one of the spatial coordinates 
\bea\label{appsol2}
&&ds^2_{BH}=-f dt^2 +dy^2  +{ dr^2\over f}  + r^2 d\Omega_{(D-3)}^2 ,
\eea 
The function $f(r)=1-{r_0^{D-4}/r^{D-4}}$. 
This can also be called an uncharged black string and it has
a asymptotic  geometry $R^{D-1} \times R^1$. 

Let us make the following boost  along  $y$,
\bea 
dt\to \gamma dt +v \gamma dy, ~~~
dy\to \gamma dy +v \gamma dt
\eea
so we get a 
 boosted BH geometry 
\bea\label{appsol2de}
&&ds^2=  
(1+{r_0^{D-4}v^2\g^2\over r^{D-4}})dy^2-
(1-{r_0^{D-4}\g^2\over r^{D-4}})dt^2+
{2 v\g^2 r_0^{D-4}\over r^{D-4}}dt dy
+{dr^2\over f}  + r^2 d\Omega_{(D-3)}^2 \br
\eea 
We  note that the periodicity of Euclidean time in the 
boosted geometry gets modified and it is  now 
${4\pi r_0\gamma\over D-4}$. 
Correspondigly  horizon temperature after the boost is
\be
T= {D-4\over 4\pi r_0\gamma}. 
\ee
Thus these BH solutions have two parameters $r_0,\gamma$.

\ud{\it The simultaneous $r_0\to 0,~\g\to\infty$ limit:}
We incorporate the double limit 
such that
\be\label{appsol21}
r_0\to 0,~~~~ \g\to\infty, ~~~~r_0^{D-4}\g^2=\beta^2= {\rm fixed}.
\ee
In which case the BH temperature would behave as 
\be
T\sim O( r_0^{D-6\over 2})\sim 0, ~~~{\rm for}~~~D> 6
 \ee
However such double limits exist  only for  
$D> 6$. For $D=5$ the temperature would blow up under the limit $r_0\to 0$, 
so we do not include them here.  
Under the double  limits we  get following wave like solution 
(introducing light-cone coordinates)
\bea\label{appsol2de1}
&&ds^2= {\beta^2\over r^{D-4}} (dx^{-})^2 -dx^{+}dx^{-}  
+{dr^2}  + r^2d\Omega_{(D-3)}^2, ~~{\rm for }~~D>6 
\eea 
We can notice that light-cone time is null, i.e. $g_{++}=0$,
in  these  wave solutions.  
These are the simplest type of plane-wave solutions 
in flat spacetime \cite{horo}. 
 Actually the expression for $g_{--}$
in Eq.\eqn{appsol2de1} is nothing but a harmonic function over 
the transverse $(D-2)$-dimensional space. 

Next  we can consider a  `bubble' like spatial geometry 
(Euclidean Schwarzschild type geometry) in $D>6$ spacetime dimensions
\bea\label{appsol2a}
&&ds^2_{bubble}=- dt^2 +g dy^2  +{ dr^2\over g}  + r^2 d\Omega_{(D-3)}^2 ,
\eea 
The function $g(r)=1-{r_b^{D-4}/r^{D-4}}$, with coordinate
range $r_b\le r \le\infty$. 
The asymptotic spacetime geometry is $R^1\times S^1\times R^{(D-2)}$, 
having a $y$ periodicity
$y\sim y+ {4\pi r_b\over D-4}$. The asymptotic 
periodicity of $y$ is chosen such that there 
is no singularity at $r=r_b$. Now
we make a boost along $y$ 
\bea\label{appsol2d1}
&&ds^2=  -(1+{r_b^{D-4}v^2\g^2\over r^{D-4}})dt^2+
(1-{r_b^{D-4}\g^2\over r^{D-4}})dy^2+
{2 v\g^2 r_b^{D-4}\over r^{D-4}}dt dy
+{dr^2\over f}  + r^2 d\Omega_{(D-3)}^2\br 
\eea 
Note that the periodicity of $y$ in the 
boosted  geometry gets modified to  
$y\sim y+ {4\pi r_b\gamma\over D-4}$. 
Now consider the double  limits
\be\label{appsol21a}
r_b\to 0,~~~~ \g\to\infty, ~~~~r_b^{D-4}\g^2=\beta^2= {\rm fixed}.
\ee
Under this limit $y$ circle essentially gets decompactified, if we keep $D>6$.
 It  gives another wave geometry 
\bea\label{appsol2de8}
&&ds^2_{}= -{\beta^2\over r^{D-4}} (dx^{+})^2 -dx^{+}dx^{-}  
+{dr^2}  + r^2d\Omega_{(D-3)}^2 
\eea 
which is also a wave in empty space but has nontrivial $g_{++}$ component.
Thus our double limits 
generally produce wave like solutions.

\section{Kaigorodov  Spaces}

Let us consider the following 5-dimensional metric
which is a solution of the Einstein gravity 
with a negative cosmological constant,
\be
ds^2_{Kaigorodov}= \pm{\beta^2 z^2} 
(dx^{\mp})^2 +{-dx^{+}dx^{-}+ dx_{1}^2+dx_2^2\over z^2}  
+{dz^2\over  z^2}
\ee
where both plus and minus signs can be considered. In our conventions
 the solution with upper signs 
 would classify as Lifshitz like, while with the lower
signs we will identify it as Schr\"odinger spacetime.
The Riemann tensor for this metric can be separated as 
\be
R_{\mu\nu\lambda\rho}=
(g_{\mu\lambda}g_{\nu\rho}-g_{\mu\rho}g_{\nu\lambda}) 
+\delta_{\mu\nu\lambda\rho}
\ee
The expression $\delta_{\mu\nu\lambda\rho}$ collects deviations from 
pure AdS geometry (the $\beta$-dependent
 terms). 
For maximally symmetric Einstein spacetimes such as AdS, 
these extra terms are of course vanishing. 
But for  above Kaigorodov-Lifshitz metrics (upper signs)
 the nonvanishing components are
\be
\delta_{-1-1}=2\beta^2,~~~
\delta_{-2-2}=2\beta^2,~~~
\delta_{-z-z}=-4\beta^2,
\ee
While  for   Kaigorodov-Schroedinger metrics (lower signs)
 the nonvanishing components are
\be
\delta_{+1+1}=-2\beta^2,~~~
\delta_{+2+2}=-2\beta^2,~~~
\delta_{+z+z}=4\beta^2\ .
\ee
Due to these contributions the corresponding Weyl tensor
is  nontrivial. 
As a consequence the Kaigorodov spacetimes are not conformally flat.
 But
interestingly these corrections to the Riemann tensor 
do not contribute to the Ricci tensor. 
The Ricci tensor for Kaigorodov Lifshitz/Schroedinger 
spaces is still given as
$$R_{\mu\nu}=-4 g_{\mu\nu}$$ 
just like for the pure $AdS_5$ spacetime.
For the anti de Sitter spacetimes, which are also conformally flat,
 Weyl tensor is vanishing. More discussion about Kaigorodov spacetimes 
can be found in \cite{clp}.
 }


\end{document}